# Ethical considerations in an online community: the balancing act

Cécile Paris[1], Nathalie Colineau[1], Surya Nepal[1], Sanat Bista[1] and Gina Beschorner[2]

## Abstract

With the emergence and rapid growth of Social Media, a number of government departments in several countries have embraced Social Media as a privilege channel to interact with their constituency. We are exploring, in collaboration with the Australian Department of Human Services, the possibility to exploit the potential of social networks to support specific groups of citizens. To this end, we have developed Next Step, an online community to help people currently receiving welfare payments find a job and become financially self-sufficient. In this paper, we explore some ethical issues that arise when governments engage directly with citizens, in particular with communities in difficult situations, and when researchers are involved. We describe some of the challenges we faced and how we addressed them. Our work highlights the complexity of the problem, when an online community involves a government department and a welfare recipient group with a dependency relationship with that department. It becomes a balancing act, with the need to ensure privacy of the community members whilst still fulfilling the government's legal responsibilities. While difficult, these issues must be addressed if governments are to engage with their citizens using Social Media.

## Introduction

Figures from recent surveys indicate that a growing number of people spend an increasing amount of time online using social media (SM) technology. According to a recent Sensis report on social media (Sensis, 2011), 62% of online Australians have a presence on social networking sites, while a recent Pew Internet survey (Madden and Zickuhr, 2011) reported that 65% of online American adults use social networking sites. These figures are remarkable if we consider that, during the period of 2005-2011, the percentage of online American adults using social networking sites went from 8% to 65%.

Capitalising on the opportunities SM brings, businesses have started building online consumer relationships, and a number of government departments in several countries have embraced social media as a new channel to interact with their constituency. Governments at all levels (i.e., federal, state and local) are taking a more collaborative approach in their dealing with citizens through consultation and participation with the aim to improve the quality and responsiveness of government policy making and service delivery (e.g., Colineau *et al.*, 2011b; Toland, 2011; Howard, 2012).

---

[1] CSIRO ICT Centre, Australia --  *email:* [FirstName.LastName@csiro.au](FirstName.LastName@csiro.au)

[2] Department of Human Services, Australia --  *email: gina.beschorner@humanservices.gov.au*

However, there is still a long way to go. For many government departments, SM tools remain fairly new, and SM communication strategies still need to be put in place. There is a need also to educate public servants on how to use these tools in the context of their work and establish SM policies and guidelines to empower them to engage meaningfully online (Hrdinova *et al.*, 2010). These new policies bring issues related to both employee and citizen conduct, security and legal issues, and therefore must be framed within existing government guidelines, including public servant values and code of conduct, privacy law, accessibility guidelines, to name a few.

In this context, through a twelve month trial, we are exploring, in collaboration with the Australian Department of Human Services (thereafter referred to as Human Services), the possibility to exploit the potential of social networks to support specific groups of citizens. To this end, we have developed Next Step, an online community to help people currently receiving welfare payments find a job and become financially self-sufficient. The purpose of this social networking site is to complement existing welfare transition programmes, providing an additional channel to deliver social services. For the researchers on the project, the site provides real data to study the development of an online community and various behavioural aspects at play in such environments.

While an instance of e-government, Next Step has a number of interesting and challenging characteristics making it different from other e-government initiatives and raising a number of ethical issues, from the perspectives of both the government and online research. First, in Next Step, there is a power relationship, as our community members are dependent on Human Services for their payments. Assurance of anonymity in this context is paramount, if we want the community members to feel free to express themselves without fear of retribution. Second, Next Step does not exist solely to seek input from the public, but rather to attempt to create a space where people can access tailored information and be linked in with relevant support services. Finally, we know from our initial research with the members of the target group that they have serious concerns about their situation and that many felt anxious and overwhelmed (Colineau *et al.*, 2011 a & b). The intent of the community is also to give people in this target group a sense of community, with opportunities to share experiences and support one another.

Ethical issues thus include anonymity, privacy and confidentiality of the information shared, legal issues as proper guidelines need to be put in place for people to be treated not only fairly and objectively, but also respectfully. Furthermore, care must be taken to prohibit unlawful and offensive material.

In this paper, using the online community we developed as a case study, we explore some ethical issues that arise when governments engage directly with citizens, in particular with people on welfare payments, and when researchers are involved. We examine some of the issues that have been raised by others in the context of online research and discuss some of the challenges we faced and how we addressed them.

# The *Next Step* online community – The parties involved

We first need to describe our context in more detail. The Next Step online community was developed to support welfare recipients, in particular parents currently in receipt of parenting payments and who now are transitioning to a new income support benefit, with the requirement to find a job. This transition happens when their youngest child reaches school age.

As welfare programmes have changed over the years, financial assistance is now provided in exchange of work (or some form of community participation) and offered for a limited period of time. The transition back to work can be difficult for some parents, in particular for single parents and people who have been out of the work force for several years. Through the use of SM technology, Human Services wants to reach these specific communities by providing them with a channel to help them develop skills and a support network that can assist them during this transition period.

In this context, our aim is to explore whether this medium can help this group in several ways:

- To be better equipped to find a job, through a reflection journey designed specifically for the community (Colineau *et al.*, 2013).
- To feel more supported in their transition, emotionally, through connections with others and the sharing of experiences, and with respect to the information they can access, by receiving information tailored to their needs (Bista *et al.*, 2012b).
- To provide a channel through which they have access to experts.

This project includes three parties, as illustrated in Figure 1. Besides the community members, the project brings together two organisations with different roles and responsibilities, and differing ethical considerations.

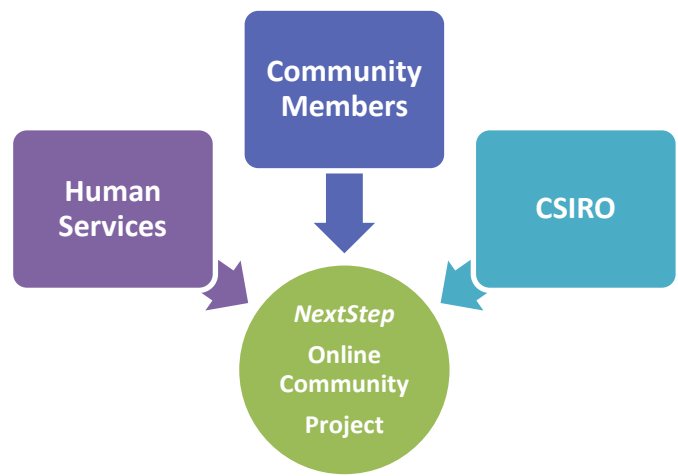

Figure 1. Parties involved in Next Step

On the one hand, there is Human Services. The people we are targeting (i.e., the community members) are amongst its "customers". Human Services is responsible for their payments and for informing them of the transition process (i.e., what to do and what payments people can claim considering their specific and individual circumstances). It is also worth mentioning that Human Services has a "duty of care" to its customers. This includes providing a service that takes people's individual circumstances into account and providing intensive services to people at a time when they need it. The Social Security Law with which Human Services must comply states that its staff must take a reasonable degree of care when providing advice and that carelessness or negligence can be a breach of this duty of care.

Human Services' interest in this project is to explore new ways to support citizens and deliver its services, and to gain insights into this specific group. Its specific role in the community is to ensure accurate information is being disseminated, answer questions that arise, provide experts when required, and, because of its duty of care, moderate discussions in the community, ensuring there is no abuse, no disclosure of sensitive information and any mentions of customers indicating life danger are immediately given support. To carry out this project, Human Services had to go through its internal processes of approval for privacy and legal issues. The Privacy Act creates a set of legal obligations for agencies such as Human Services in terms of their handling of records containing personal information. There are also community expectations regarding the government's handling of records containing such information. Human Services carried out a comprehensive Privacy Impact Assessment to assist in identifying and properly managing the 'privacy issues' associated with the Next Step project and avoid a breach of the Privacy Act. The assessment was undertaken by a third party organisation and coordinated by Human Services' privacy team. The report contained eleven recommendations to improve the handling of customer information.  These included: measures to remind community members that information posted in the community was being collected; safeguards for members' personal information; a requirement for a random audit at least every three months to ensure that records containing personal information were being securely kept; a feedback form to be provided in the community permitting members to requests amendment or deletion of personal information contained in posts; and the instruction to delete, at Human Services, all electronic material related to Next Step in 2017. All recommendations were accepted and implemented.

In addition, the Legal Services Branch developed Terms and Conditions to inform customers of the community rules and adequately protect Human Services. They also contributed to, and approved, the 'Moderator Guidelines' which advise when and how Human Services moderators should respond to customer comments. These guidelines clearly state that "The moderator's primary focus is to follow discussions, promote conversation and moderate any breaches on the forum. The role of the moderator is to assist in building the community, ensure the safety of members, and ensure compliance with the site's Terms of Use. Moderators also help keep members focussed on important issues by policing spam and irrelevant discussion threads or posts. An example of a specific guideline calls for moderators to "watch for offensive words and for breaches of the Terms of Use".

On the other hand, IT and Human Computer Interaction (HCI) researchers are also present. They are researchers at CSIRO (Commonwealth Scientific and Industrial Research Organisation), the national research institute in Australia. Their aim in this project is, as researchers, to study a number of issues related to online communities. This includes research questions on how to develop a sense of community, foster interactions and create a sustainable community; on how to create and develop social trust, and on how to encourage reflection to induce behavioural and attitudinal changes (Nepal *et al.*, 2011; Colineau *et al.*, 2013).

The researchers designed the online community portal with the help of Human Services. The site was developed and is hosted by CSIRO. The researchers are responsible for the running of the community portal. They are also responsible for studying the community activities, collecting and analysing data to research the questions mentioned above. To do so, they developed and introduced a number of mechanisms – e.g., gamification (Bista *et al.*, 2012a), buddy matching programme (Colineau, 2012); activities (Colineau *et al.*, 2013), etc. – aimed at encouraging interactions amongst community members.

Before the trial could start and for research to proceed, CSIRO researchers had to follow an internal process of approvals, including the ethical clearance, following the CSIRO's Human Research Ethics Policy which complies with the values and principles specified in the National Statement on Ethical Conduct in Human Research (2007). The ethics process ensured that the methodological and analytical framework used to analyse and measure the utility of the social network site was clear. Particular attention was given to providing explanations about the technical steps put in place to protect the privacy of the personal information collected throughout the trial. A clear and consistent approach to managing the dual missions of Human Services was put in place, in particular, (a) with respect to Human Services' role in providing informational and online moderation support, and (b) with respect to the re-identification of data to ensure it will not be sought by Human Services unless justified by potential risks of harm to participants. Information provided to participants (including the terms and conditions statement) had to be provided in plain English and included a clear statement that any information provided will only be identifiable in specific circumstances (described later in the article).

The third party in the equation is of course the invited participants, the community members. They were recruited directly by Human Services through their secure online mail service. Only people who fitted the target group, had signed up for the online service and who had agreed to participate in research were contacted. We followed a double-blind process to preserve the participants' anonymity. (This is explained in more details below.)

## The Ethics of Online Research

With the growth of the Internet, and, in particular, with the surge in usage of social networking sites, people are increasingly interacting online, discussing and sharing experiences, and, sometimes, very personal information. This has opened the door to a new kind of research where data can be collected online, from a large pool of people far from the constraints of a laboratory setting. This has also raised a number of questions with respect to methodology and its associated ethical issues. The main concerns usually raised are around the "*notions of privacy and confidentiality, informed consent and narrative appropriation*" (Brownlow and O'Dell, 2002; p.686). Should the data generated online and contributed by various authors be considered private or as part of the public domain? If this kind of data is to be collected and analysed for research purposes, what are the researcher's responsibilities and obligations?

A number of researchers have discussed these issues more generally (e.g., Elgesem, 2002; Weeden, 2012) or in the context of their own work (e.g., Brownlow and O'Dell, 2002). Others have proposed a set of guidelines: for example, Sharf (1999) derived hers from her work with online breast cancer discussion groups; Elgesem (2002) discussed the Norwegian ethical norms and its relevance to Internet research, and McCleary (2007) provided a set of recommendations based on the ethical principles of the Belmont Report (National Commission for the Protection of Human Subjects of Biomedical and Behavioral Research, 1979) and the NASW Code of Ethics (NASW, 1996). However, as noted by Weeden (2012), whereas the issues of Ethics arising from online research have been given a great deal of attention, there is still a need for clear standards and guidelines as legislations and requirements are not consistent across boards and can change from one Institutional Review Board to another.

Coming from an IT perspective and looking at information as an individual's intellectual capital, Mason (1986) focused on four particular ethical issues related to the sharing, collection and distribution of information about individuals or generated by individuals. He discussed in particular the ethics implications of information *privacy* (i.e., the disclosure of one's personal information and/or one's association), *accuracy* (i.e., the fidelity of the information being disclosed, collected and kept as records), *property* (i.e., the protection of intellectual property rights through various mechanisms) and *accessibility* (i.e., the understanding of information and the means by which people gain access to it).

In line with these considerations, we now discuss the ethical issues for Next Step. We argue that ethical issues should be of concern not only to researchers doing online research but also to governments wanting to engage with their citizenry via social media.

## Ethical issues for Next Step

We first focus on Mason's (1986) four key ethical concerns. We then turn our attention to some of issues raised by Sharf (1999) and discussed by Brownlow & O'Dell (2012), namely the responsibility of the researchers – in our case, both the CSIRO researchers and staff from Human Services.

In the context of *Next Step*, we consider a specific set of questions for each of privacy, accuracy, property and accessibility as follows:

- Privacy: What information about one's self must a person reveal to others, under what conditions and with what safeguards (confidentiality issue)? Can people keep things to themselves? If yes, what?
- Accuracy: Who is responsible for the accuracy of information disseminated in the community?
- Property: Who owns what information? Who owns the channels through which information is transmitted?
- Accessibility: We address here the issue of fairness of access and citizen equality, which is important when dealing with welfare recipients.

We review below how these issues play out in our context.

**Privacy (& Confidentiality)**

Privacy is related to what information can be revealed or shared with whom. There are a number of privacy issues in our community. This includes, for example, the identity of the participants; how much information can a member reveal to other community members, to the CSIRO, i.e., the researchers? Should it be kept confidential, to whom in particular and under what circumstances?

This also includes the information shared with or about the community members. For example, do we provide community members a summary of the data collected? How much information can or should CSIRO reveal to Human Services while reporting the results/outcomes?

We have addressed (and to some extent are still addressing) privacy and confidentiality issues at different phases of the project, sometimes in different ways to suit a particular context: for example, during the recruitment process, while the community is running and when reporting about the community.

*Confidentiality of members' identities*

Privacy and confidentiality issues had to be addressed during the recruitment process. We use a double blind recruitment protocol, which is shown in Figure 2. It works as follows: (a) CSIRO creates tokens which are provided to Human Services; (b) Human Services uses these tokens to send an invitation letter to the targeted group of participants, selecting them from their customer database; (c) Participants register using their token, and select in the process a screen name to represent themselves in the community. This information flow is indicated by the thick green arrows between the three parties in the figure. With this process, Human Services knows the relationship between the tokens and the customers' identity. CSIRO knows the relationship between the tokens and the screen names. Neither CSIRO nor Human Services know the actual identity of the community members. Once in the community, the screen name (the name members choose for themselves in the community) and content posted on the forum is visible to everyone. What is known (visible) to whom is illustrated with the blue circle and the thin blue arrows at the bottom of the figures. This process guarantees the anonymity of the community members. Similarly, the content generated by members is associated with their screen name; therefore Human Services cannot link any content with particular individuals. Without knowing the token-screen name-customer relationship, members' data cannot be re-identified. In this scenario, CSIRO acts as an impartial "trusted third party", one with no relationship with the community members, no access to members' customer records within Human Services, and who guarantees that it will keep the confidentiality of the information bestowed to it. We thus created a privacy triangle, with CSIRO acting as the trusted third party, as illustrated in Figure 3.

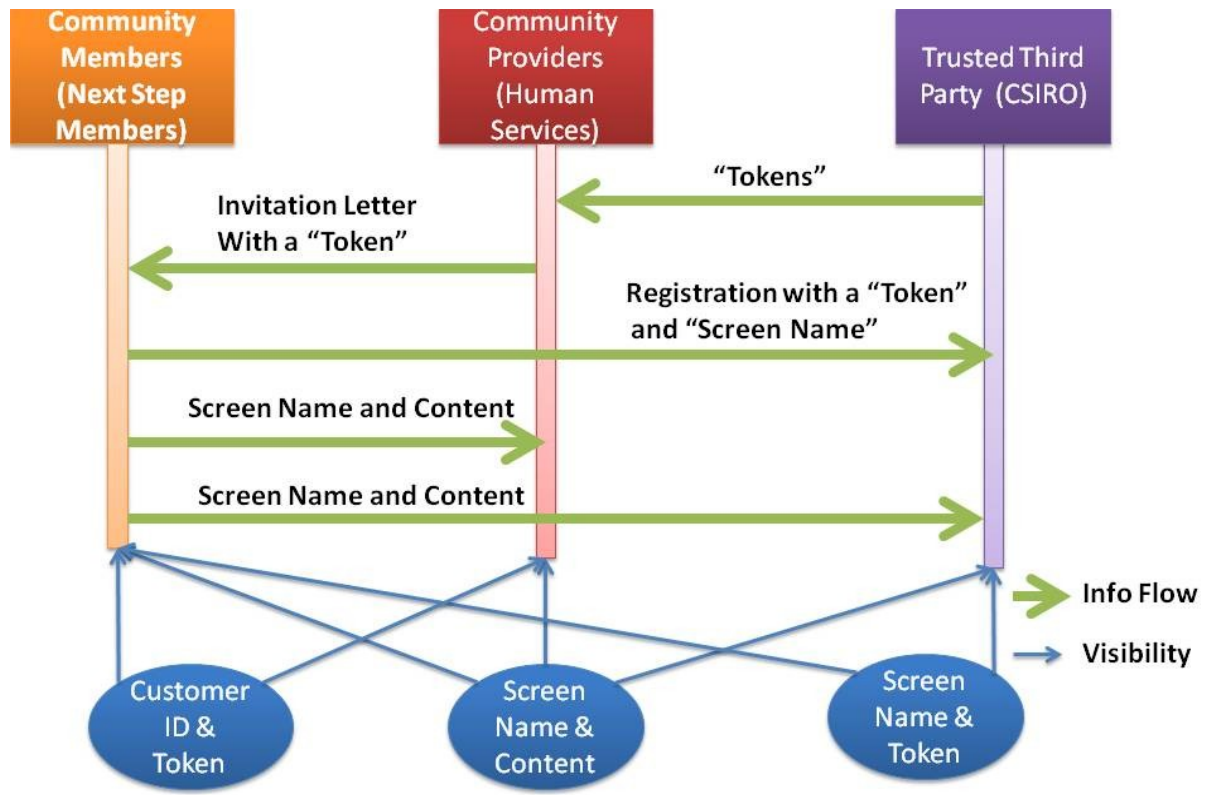

Figure 2. Recruitment and Interaction Protocol

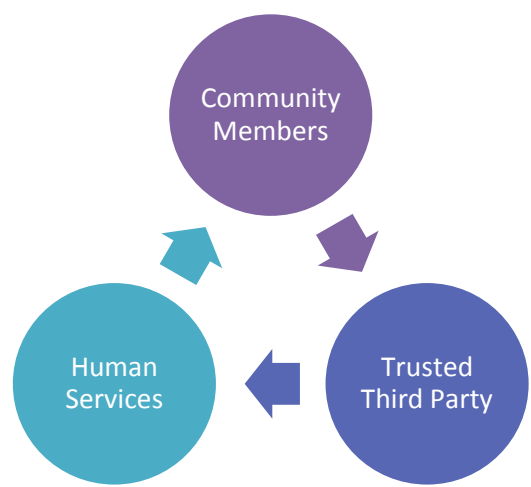

Figure 3. The Privacy Triangle. Ensuring anonymity: CSIRO sends token to Human Services, who invite members; these sign into the CSIRO platform with their token

This process is important given the dependent relationship that exists between community members and Human Services. If the CSIRO researchers were not in this privacy triangle, we argue that a trusted third party would be required instead, to ensure participants' anonymity with respect to Human Services. This trusted third party would also be the owner of the platform. At the onset of the project, prospective members were told their anonymity would be preserved, and that Human Services would not be able to link them to their customer record, except in special and extreme circumstances described below, to fulfil Human Services' responsibilities.

*Respecting privacy and confidentiality while fulfilling ethical and legal responsibilities*

The ethical issue around privacy (and confidentiality) is not a simple one when it comes to our community for two reasons. First, we have a responsibility of *duty of care* towards community members. This is true both for the Human Services staff (towards their customers) and for CSIRO researchers conducting research involving human participants: In case someone shows signs of distress, we are obliged to intervene to ensure the wellbeing of the concerned individual. This required us (i.e., CSIRO and Human Services) to agree on a contingency plan in case we needed to intervene. This included in particular discussing what to do and clarifying who would intervene. The emergency response guidelines ask the moderator who notices the situation to call their team leader to discuss the comment; the team leader may in turn contact their National Manager to decide whether they need approval to re-identify the customer.

Partially to address the duty of care, conversations in the community are continuously monitored. When a distress scenario is identified, the level of distress is assessed. If it is not deemed to be life threatening, the individual member is invited to contact CSIRO, via a special Next Step email address created for that purpose, the CSIRO Ethics Officer, or Human Services directly (the phone number of a Human Services staff trained to handle such situation is provided). It is then at the discretion of the individual member to make such contact. If the distress is deemed life threatening, the CSIRO researchers need to provide Human Services with the token of the participant in order for a trained staff member to intervene. We developed, together with the ethics committee, guidelines to decide when the situation warranted identity disclosure. In the past six months, we have experienced only one case in which we felt that a person was distressed. The person was invited to contact someone at Human Services directly, or to let us know if he or she wanted his or her identity to be revealed to Human Services for them to contact him or her. We have not had an instance in which the CSIRO researchers had to reveal someone's identity to Human Services. We believe guidelines akin to the ones we have established should also be in place if governments were to run an online community with specific target groups such as welfare recipients, without researchers like CSIRO staff involved acting as a trusted third party.

Second, Human Services has the *legal obligations and responsibilities* of any government agency to report instances of fraud if made aware of any (i.e., Human Services' staff in our community do not actively look for fraudulent behaviour, but, if illegal behaviour is explicitly mentioned or apparent, they are obliged to report it). Duties of care and legal obligations thus have a direct impact on privacy and confidentiality issues since they entail that Human Services has access to the identity and/or contact details of the individual in question. This is against our original guarantee of anonymity!

We address these issues as follows: first, community members are explicitly told about the special conditions under which their contact details will need to be given to Human Services ("*Please note that, under some special circumstances, Centrelink will have the ability to identify and contact participants if their welfare is at risk or if they have disclosed some illegal activities.*")[3]. This is explained in the participant information sheet accompanying the invitation letter, in the consent statement participants have to agree to and again repeated in an information page about the community, always present on the site. It is worth noting that, as in the case of distress, the disclosure of information potentially related to illegal activity is also assessed individually. For example, the following comment was made in the community forum:

---

[3] At the start of our project, the agency responsible for this target group was called Centrelink. It has now been regrouped with other agencies such as Medicare to form the Department of Human Services. Clients usually still refer to the agency as Centrelink.

> *"I'm not even telling centrelink that I am studying because then they will take me of the jobseeker list and I will lose the help that i receive from my JSA[4]. I would rather leave my options open than be forced to start from scratch again once my course is finished"*.[5]

Although not related to an illegal activity, this is referring to the non-disclosure of certain information to Human Services. Human Services moderators reviewed the comment and determined that, as no illegal activity was taking place, they would not re-identify the individual - instead, they drafted a response to remind the person of the importance of having their details up to date with Centrelink to ensure they receive all of the payments and services they are entitled to, and to avoid receiving an overpayment and incurring a debt. We have not had so far any instance of illegal activity requiring CSIRO to disclose a member's token to Human Services.

It is therefore important to ensure members understand the implications of the ethical and legal responsibilities Human Services have, as the disclosure of their identity under special circumstances may have serious consequences.

*Privacy and Confidentiality of information provided by members*

As members of the community, people provide information: in their profile, in their public discussions, as part of the activities that are offered in the community as a reflection journey, and simply as participants (i.e., logging information, etc.). One must then consider how much of this information is provided to whom for which purpose. We examine each type of information in turn.

*Profile*

Members are asked to fill in a profile, in which they reveal as much or as little as they wish. How much they reveal to others is thus totally under their control. The profile is meant to enable members to present themselves to others in the community. The profile includes several parts, which are treated differently with respect to what information is disclosed to others:

---

4 Job Services Australia, JSA, is the Australian Government employment services system that supports job seekers and employers.

[5] All quotes from the forum are *verbatim* – punctuation, syntactic or spelling mistakes are original.

- Background: in this part of the profile, participants are asked for demographic information. This is used by the researchers for their analysis. Only CSIRO researchers see this information.

- General: here, we ask participants to present themselves to others in the community, by indicating their mood and writing a few sentences about themselves, their family, their local community, their hobbies, and their dreams. The page clearly tells participants that everyone will be able to see it, including Human Services staff.

- Skills and Attributes: in this part of the profile, we ask participants to think about their skills (and weaknesses) and their qualities. Participants have the choice to make that information private (i.e., no one can see it), general (i.e., everyone can see it), or visible by their friends only (i.e., people in the community with whom they have become "friends", or "buddies"). This is illustrated in Figure 4.

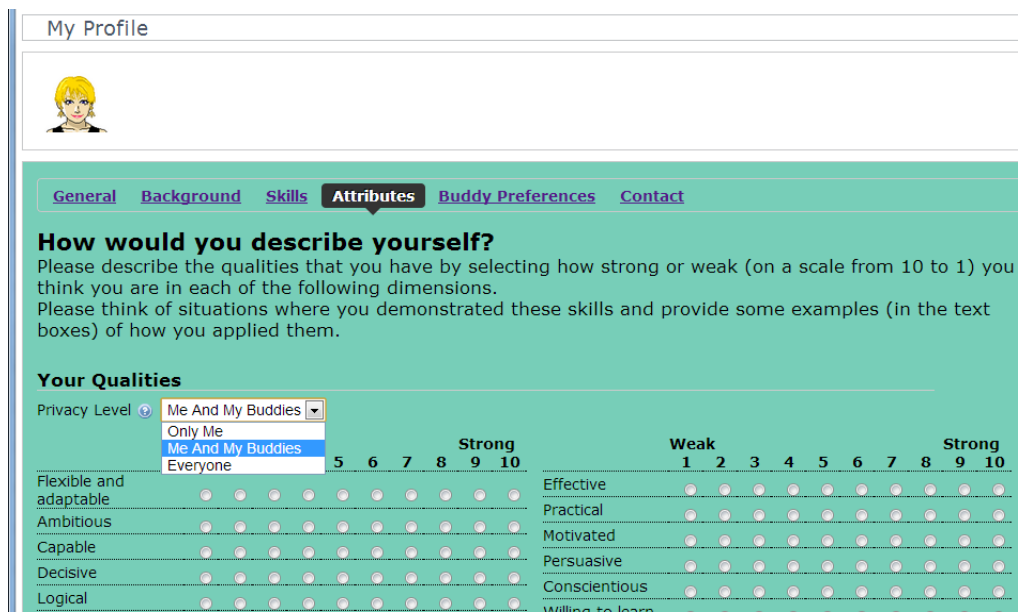

Figure 4. Portion of the screen for the user profile: it indicates how the participant can choose who sees the information

*Discussions*

All forum discussions are "public", meaning they can be seen by the whole community. Members know that what they write can be read by everyone. As Human Services staff and CSIRO staff often start discussion topics and, in some cases, answer questions, it is clear that they read what is in the forum. The following posts provide two examples of posts from a Human Services staff member (Gigi-from-human-services): one as a starting post for a new discussion, the other as a reply to a member's post:

*"Hi everyone,*

*Hope you're all having a safe and pleasant holiday season. Just thought I'd let you all know we have added a new resource in the toolkit on setting goals in the new year. You can check it out here. We'll be adding more resources soon, so make sure to check out the toolkit section. Let us know if there is any info you want us to share, whether it's payments and services related, or more related to lifestyle topics."*

and:
*"Hi Lillypilli,*
*i have double checked this for you and found out that […]"*

It is interesting to note, in fact, that, in the requirement analysis we performed prior to the development of the community, through a number of group interviews and an online survey (Colineau *et al.*, 2011a and b), prospective participants clearly indicated that they wanted the forum discussions to be monitored, both to ensure the accuracy of the information and to avoid discussions turning into "*purely griping sessions*".

It is also clear from member' posts that they are aware that Human Services staff read the forum, as illustrated by posts such as: "*I was hoping to hear from Gigi, i thought she may able to tell me*", or posts directly addressed to Human Services staff (e.g., *"[…]I really appreciate you finding all tis info for me Gigi so I can be prepared next week"*; *"Thank you, Marian! I can't tell you how much of a relief that is to me!"*, *"Hi Gigi"*, or *"Thanks Gigi, your support is great"*.

Given the confidentiality we have guaranteed to members, we hoped people would discuss issues freely. Our experience so far is that people have indeed spoken their mind. For example, the forum has had comments such as *"… not that Centrelink cares" ; "I just wanted to let other members here know to maybe not rely on Centrelink informing them of anything…"; "NO I DON'T FEEL SUPPORTED !!",* even though participants know that Human Services staff are reading the posts. We also feel that people are interacting naturally, given comments such as *"hi Kayte, would you like to have a chat, I understand your* *frustration, I know a little about the changes, maybe two heads could work on your situation with your health card."*

*Activities*

We provide activities to participants, taking them on a reflection journey to help them make the transition to work (see Figure 5) (Colineau *et al.*, 2013). This journey is meant to guide parents through a series of steps to help them face obstacles, regain confidence and plan their return to work. When someone participates in an activity, he or she typically is asked to provide information. For example, an activity might ask participants to think about their personal achievements and then list them in a table provided to this effect, together with an explanation as to why they consider it an achievement. The information provided in the activity is private, when the activity is to be done individually, or shared with friends when it is to be done in collaboration with others in the community. The CSIRO researchers have access to the information, not Human Services.

Figure 5. A reflection journey through activities

*Information from logging, browsing, etc.*

CSIRO researchers monitor all activities in the community, including when people visit the community, what discussion they read, which resources they examine, etc. This is to enable the researchers to perform a number of analyses on the community, and provide feedback to Human Services as to the popularity of the resources or discussion topics. Based on their analysis, the researchers provide regular reports to Human Services. No individual confidential information is included in these reports (e.g., no information from the demographic part of the profiles), and it will not be included in research reports that might be published. Similarly, as activities are meant to be a personal reflection, individual information contained in the activities is not provided to Human Services staff.

Based on behavioural information, we have introduced badges, which are equivalent to loyalty points for various actions in the community: for example, the reader badge is given to the community members who read and rate the most number of resources/posts. Badges give us a quick overview of the state of the community and peoples' behaviour (Bista *et al.*, 2012a). Individual members see their own badges. For privacy reasons, we currently do not make these badges public to the rest of the community. We have asked community members whether they would like to make this information public (i.e., know that member X is reading a lot). We are waiting for answers on these questions to decide whether we should change our current settings. CSIRO researchers see all the badges, and make them available to Human Services as well. We do not feel that revealing badge information to Human Services is problematic ethically, as they do not reveal private information.

The Next Step community is hosted on a secure site at CSIRO, and all the data collected is stored on secure computers also at CSIRO. Brownlow and O'Dell (2012; p 687) point out that "researchers need to be careful about the assurance they give to participants regarding confidentiality", given that it is hard to guarantee the security of electronic communication. Recognising this problem, we explicitly told our participants in our terms and conditions that "*No data transmitted over the Internet can be guaranteed as totally secure*", although "*we strive to protect your communications*". In agreement with Brownlow and O'Dell, we believe this should be made explicit to participants.

In terms of fidelity of the information provided by participants, all information is saved verbatim in a secured data base. In our analysis, we will do our best to ensure we represent the views and stories shared by community members accurately, when aggregating the data.

**Accuracy**

Human Services is responsible for the accuracy of the information provided within the community, for example in the resource section or in answer to questions in the forum. The Human Services staff involved in the community are from the communication division. They follow an established internal clearance process when responding to customers in the forums to ensure information is approved, typically consulting back with the specific business teams. The clearance process ensures that all information that is not publically available is approved by the appropriate business area, at the necessary level, before being published. Generally responses are published the same day or within 24 hours. These clearance processes also apply to other content such as resources, videos and podcasts. Experts in the live chat sessions, such as social workers or policy experts, have the authority to answer questions in their field of expertise in real time. If they are uncertain of an answer during a live chat, moderators advise the community member that their question will be followed up and an answer provided in the forum. The response to the question then follows the clearance process described above.

All Human Services moderators have been trained on the legal and privacy issues that relate to social media moderation. To support them in their role, there are 'Moderator Guidelines', already mentioned earlier in this article. These define the role of Human Services moderators, the rules of the community, what constitute a breach of the Terms of Use and the actions to take, other issues to monitor, and escalation policy. The Human Services moderators work full time in online community moderation and take quick action when there is a need. Moderators are objective and impartial ─ they do not censor or delete negative comments, only those that could threaten or disrespect other community members (there has not been a need for moderators to delete or amend any comments thus far). We assume that stories related by participants are true, at least from the participants' perspectives.

**Property**

Human Services owns the information related to their processes, information sheets, and the video and podcast resources they produce to publish in the community.

Participants were informed in the Terms of Use of the community that they assigned to the CSIRO all copyrights contained in their communication on the site, and that CSIRO researchers could use it (essentially for research and publication purposes). The CSIRO thus owns that information, as well as all the data collected. Again, this was specified in the Terms of Use and consent form.

We have set out specific guidelines to deal with the data collected that are as follows:

- At the end of the trial, a summary of the results will be sent to all participants who chose to receive it. In addition, the work will be disseminated to the public through publications to journals and conferences. However, privacy and confidentially of members will be preserved in this process.
- CSIRO retains the data for five years for the purpose of research. After the expiration of the retention period, all records will be destroyed by commercial confidential services. This is the procedure that CSIRO has in place to ensure the content is completely wiped out the computer disks.

If Next Step was not a research project as it currently is, the issue of property might be a more delicate one. Would Human Services own all the information provided by participants? For which purpose? What would happen to the data collected? We believe that governments engaging with their citizens in online communities would need to think about such questions, especially from an ethical point of view.

**Accessibility**

There is a tension between the provision of IT related services and the fairness of access and citizen equality. To start with, there is a potential issue with access to a computer and the internet. In addition, Next Step, like many online communities, requires people to have access to a broadband network in order to get a decent response time. We have been made aware through our interactions with our participants that some only have access to a computer (and broadband) by going through a library or a training institution. Next Step is a research project, and, as a result, we did not have to address all these issues. Accessibility is, however, an issue that governments would need to address if they were to adopt SM-based support services, at least recognising that it is not necessarily available to all and might increase inequality through the digital divide.

For Next Step, we still had a number of accessibility constraints. For example, developing content and services for government means there are restrictions in terms of IT choices: accessibility guidelines apply, and web content has to be made accessible not only to people with disabilities, but also to people with limited internet bandwidths, with older versions of web browser, etc. All Australian government website must comply with the W3C recommended Web Accessibility Guidelines 1.0 (WCAG10) by December 2013.

As Next Step is a research trial at this stage, we have not had to comply with *all* the accessibility guidelines. (It is also worth reminding the reader that we have only invited people who have already signed up to have an electronic account with Human Services and thus already interact electronically with the agency.) We did endeavour to support multiple and older versions of browsers. Conscious of the fact that broadband access is expensive and that our target group has restricted financial means, we had to be careful in the types of applications we deployed, thus also putting constraints on the look-and-feel of the site (for example, we do not use any Flash-based modules). Finally, we met the accessibility guidelines for all videos and podcasts, ensuring for example that they all had a transcript associated with them.

Another concern was to ensure that all information available to community members would also be available for the public at large, as Human Services cannot be seen to give an advantage to one subgroup. Related to this concern of fairness is that of incentives: in online communities, in order to encourage people to join or be active in the community, incentives are often offered (Burke et al., 2009). For example, in a commercial site, people might be given a voucher or loyalty points towards the purchase of a product. Providing such encouragement is particularly useful when a community is new, as there is an immediate need to attract participants to ensure the community is alive as soon as possible. This is called the "cold start" problem: at the start of the community, there are no members, and the community developers have the challenge of bringing people to the community, a process called "bootstrapping the community", by, for example, providing incentives, or making joining and participating in the community a game, where people can "win rewards" (through high participation). This is called "gamification". In Next Step, we could not give any tangible benefits (whether monetary, informational or in terms of treatment) for fairness reasons. This made the bootstrapping of the community and the application of gamification techniques to the site difficult. We can only encourage members to participate and be active by telling them that they will learn from and support each other, thus getting support and easing their transition.

## Responsibility of the researchers and results of the research

In their discussions of ethical issues for on-line research, Brownlow and O'Dell (2012) touch upon the behaviour of the researchers: are the researchers unseen observers ("lurkers"); are participants thus ignorant of the fact that they are being observed?

In Next Step, we did not join an existing community or group as unseen observers. We were the explicit creators of the community. The community was clearly branded as government sponsored, as shown in Figure 6. The information fact sheet about the community also clearly indicated that it was a research project, in collaboration with CSIRO.

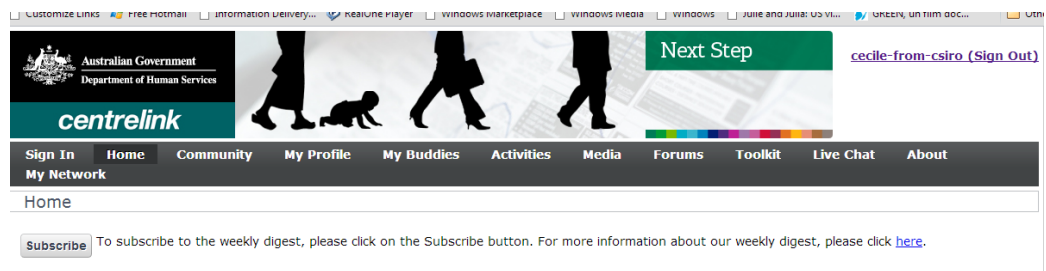

Figure 6. The Next Step Online Community, clearly branded Centrelink

As participants in the community, we are easily identifiable, as our screen names include our organisations (e.g., "*Cecile-from-CSIRO*", "*Gigi-from-Human-Services*", "*Gina-from-Human-Services*"). Human Services' staff are actively involved and present in the community, answering questions and engaging with community members on a daily basis. As already mentioned, community members often address them explicitly (e.g., "*Thank you gigi, That is excellent information.*"; "*Hi Gigi…*"; "*Thanks Marian , it's just so stressful […] Thanks for listening and replying*"; "*Thank you for clarifying that for us Gigi. :-) The link provided was very useful.*"). CSIRO researchers are less engaged in the forum discussion (because of the requirement that all information posted on the forum follow the Human Services approval process to ensure accuracy). They are still present, though, as they post resources, and their profiles are featured (see Figure 7). Community members know they are there and observing the community, as indicated both in the posts that clearly address one of them (e.g., "*Hi Nathalie…*") and in the following post: "*G'day fellow lab rats. […] it's reassuring that when I get the feeling that somebody is watching me, I know somebody actually is (I'm looking at **YOU**, CSIRO and DHS) and it's not just a case of delusional paranoia.*"). We are thus definitely not unseen observers. In this regard, we followed the guidelines of Sharf (1999) for researchers of online research to clearly introduce themselves as to identity, role, purpose and intention to the community members. The fact that the community is a research project was also made very clear to all prospective participants. It was good to see, though, that our presence and the fact that we are observing and analysing the community does not seem to inhibit community members from telling their stories or share their discontent at the process, the legislation or even Centrelink (as illustrated previously).

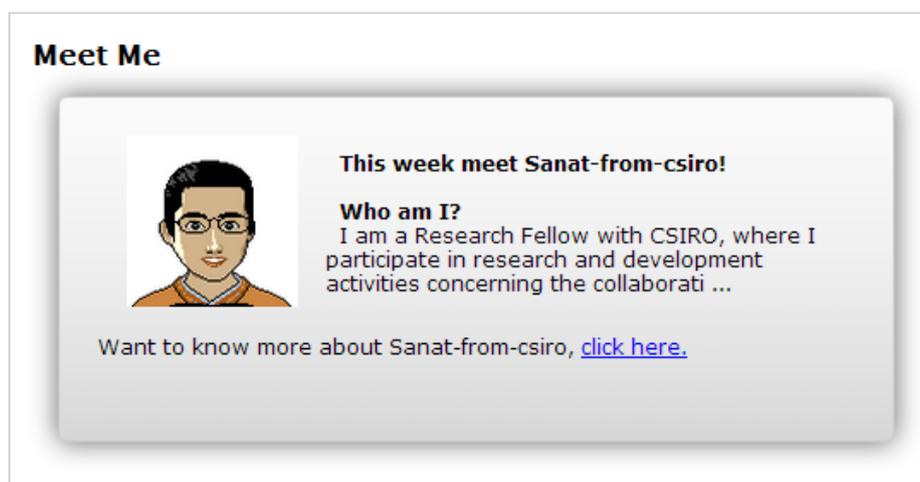

Figure 7. A featured profile from one of the CSIRO researchers

Another ethical issue that is often raised is that of the results of the research: how are they going to be used, and to whose benefits? Will the finished piece be made available to those who have participated in its creation? (e.g., Brownlow and O'Dell, 2012). In terms of research output, we told participants in the information sheet that they could choose to receive a summary of the research findings *("How can I find out more about the research? If, at the end, you would like to receive a summary of the findings of the research, please indicate so when registering your* *participation."*) As mentioned earlier, we will be sending to anyone who requested it a short paper describing the results of the research, written in plain English. We will also most likely include a link to it on the web site.

The *purposes* and *benefits* for the project are varied. The project tests the value of an online community to support the target group and enables a study of online communities and their potential in the government domain. An intended benefit was to determine whether online communities could become a new effective and efficient way to support Human Services' customers, capitalising on the potential of people helping each other and with the one-to-many communication. Through the life of the project, other benefits have emerged. Human Services gained additional insights into a specific target group, including how they understand their situation, the transition and its requirements, and obstacles they face.

For example, it became clear that participants did not have a full understanding of legislation and how things work (e.g., Human Services is not responsible for the policy relating to parents transitioning to a different payment once their youngest child reaches school age). Human Services was able to uncover some issues in existing processes and information.

For CSIRO, the benefits are around the research on a number of issues related to online communities, including their applicability and potential in the government domain. For the participants, the intended benefits included testing the possibility that they might develop a support network with others in similar situations, and get access to experts and resources to help them through the transition. We believe this benefit has eventuated. Members' posts such as the ones shown in Figure 8 (all posted by the community members, in response to other members' posts), illustrate the fact that members are giving each other support, and that this support is welcome. Posts shown in Figure 9 provide evidence that participants received information and support from Human Services that is helping them. As it turns out, an unexpected benefit to participants has been to provide them with a place to voice their views, and they have expressed satisfaction in being able to do so, and feeling that they are being heard. The latter, however, is potentially problematic. We have made it clear that, while Human Services can provide support and point people in the right direction, it is not responsible for policy. So we might end up with a mismatch of expectations among participants.

> Hello greenvanessa,
>
> I do empathise with your situation as I've been in a similar situation when I lived in a women's refuge. Though I can look back to those days, I can see that it was tiny forward steps that got me out of that situation. You have a cleaning job, which is a start - and some cleaning pays much better than it used to when I was cleaning. Next, if you have access to the internet, see if there are any jobs that can be done online - just beware of the shonky ones. Here is a freelance site which may be something to try: […]
>
> But view any job you do as a step up the ladder, and don't be too downhearted at starting with cleaning - they will always need us cleaners. In the meantime, see if you can do an online typing course to sharpen your data entry and computer skills. There used to be free computer courses for the unemployed in the days when "skill-share" was in operation - that's how I got out of cleaning and into office work. […]
>
> I hope others can give some advice to you that will be of more help. It's very hard for you now, but keep your chin up, and keep trying - we all hope that things will turn around for you soon.
>
> All the best~
>
> ————————
>
> Well if I made you laugh it was worth posting!!
>
> ————————
>
> Good Luck Emm, most cases with getting work now is not what you know but WHO you know. I wish you success in getting some temp work 😃
>
> ————————
>
> What a wonderful story Bunniesmum. Thank you for sharing and how true to life this tale can be. 😃
>
> ————————
>
> Best tip ever - find out what day your local supermarkets mark down meat and do the rounds buying it up. 2 pieces of steak for $3 is even cheaper than you can buy mince
>
> ————————
>
> Emm, I like your idea of paying the advance on your credit card, I might use some of mine to do that too and then work on my low interest loan.
>
> ————————
>
> Good info Tox!
>
> ————————
>
> Please excuse me for interrupting your message here. I noticed that you are considering work for the dole (that extra $20.80 is very tempting), and out of a sense of decency I want to give you a few pieces of information that I hope will help you. […]

Figure 8. Members' posts illustrating that members are supporting each other

> My employment pathway plan has expired and I need a new one and I dont have a job services provider because I satrted working voluntarily before i had to so I naver got one. This is why I want a face to face interview. It just seems to many things to organise in one phone call. Also who tells me if the course is approved centrelink number you provided or tafe. I really appreciate you finding all tis info for me Gigi so I can be prepared next week"
>
> ________________
>
> Thank you, Marian! I can't tell you how much of a relief that is to me!
>
> ________________
>
> Thanks Gigi, your support is great. I'd also like to thank you for your promptness in answering our queries and for looking into more serious issues
>
> ________________
>
> Thank you for clarifying that for us Gigi. :-) The link provided was very useful.
>
> ________________
>
> Thanks Gigi - I have been reading through the posts and just want to say you are doing a great job under what can only be very distressing circumstances, trying to help people when there are very few options for most.
>
> ________________
>
> Thank you gigi,
> That is excellent information.
>
> ________________
>
> Thanks Gigi, info very helpful.
>
> ________________
>
> Thanks Marian. It's good to get some positive reinforcement and encouragement. I am actually on the NEIS program. I have completed the study and am now an official business owner. I just have to get some clients now.

Figure 9. Members' posts indicating that they are obtaining useful information and support from Human Services

# Discussion

We have discussed some ethical issues in Next Step, following Mason's (1986) four key ethical concerns (privacy, accuracy, property and accessibility).

We observe that a subset of these ethical issues arise at each point in the life cycle of the Next Step community, from inception (naming and branding) to closing (who owns the data; who sees the results), as illustrated in Figure 10.

The naming of the community ("Next Step") needed special attention due to (1) the connotations it may convey to potential participants and (2) copyrights and trademarks considerations.  The branding of the community has implications to all parties involved: community provider (Human Services), CSIRO (or a trusted third party) and community members (Human Services customers).  It clearly tells members who is involved in running the community and provides the perception of ownership. In addition, a given brand often carries the brand reputation and social capital with it. It can thus play a significant role in the bootstrapping of the community. In Next Step, the branding is that of the government (Human Services). This can be both good and bad, depending on one's perspective. In any case, though, it clearly tells participants that the government is involved.

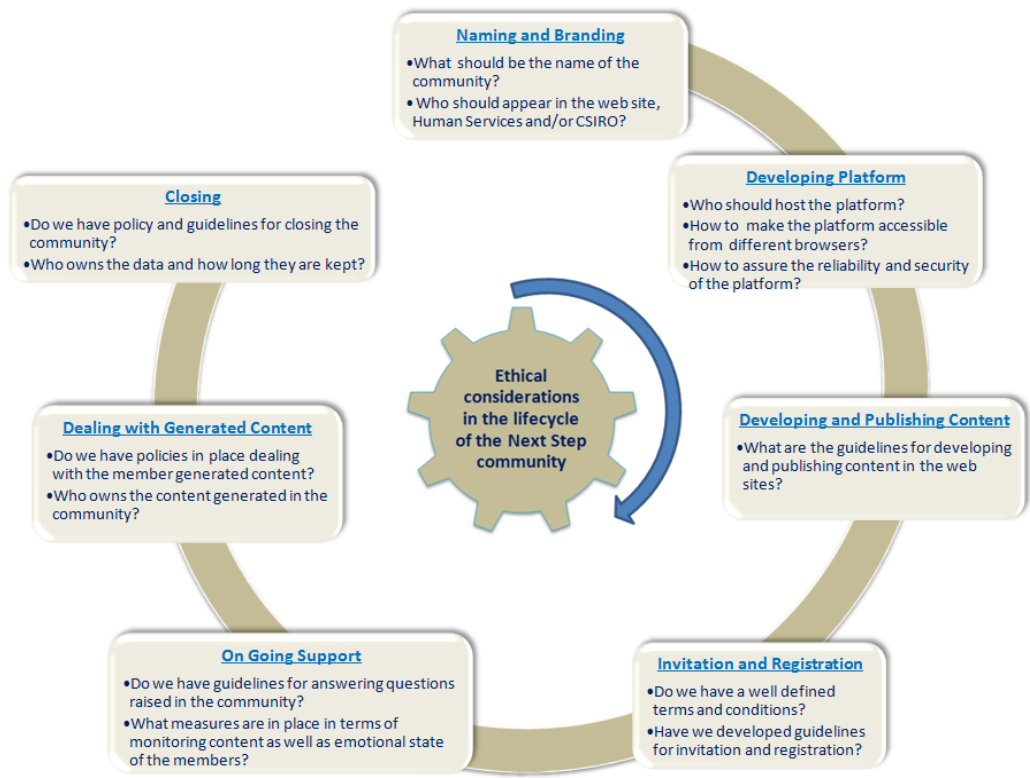

Figure 10. Ethical considerations in the lifecycle of the community

The next phase is the development of the platform. In this phase, ethical questions to be considered are: who can see what type of data, who will host the platform, how to assure the accessibility of the platform, etc. These ethical questions are addressed using the existing policies and guidelines from the two organisations: CSIRO and Human Services. CSIRO being the developer and host of the platform, its policies and guidelines play a prominent role in this phase.

Human Services is responsible for developing and publishing the content, and thus existing policies and guidelines in Human Services are applied during this phase.

The ongoing support and maintenance of the platform involves not only ethical issues related to providing accurate information in answering questions raised in forum and live chat, but also the duty of care by providing the emotional support through continuous monitoring of the community.

There are two types of user generated content. One is private, and, as a result, not visible to other community members or to Human Services. The public content is visible to all: community members, CSIRO and Human Services. CSIRO owns the user generated data.

Finally, the closing phase of the community should have explicit guidelines for dealing with data collected during the lifetime of the community: how long the data is retained, how the data is destroyed after the expiration of retention period, etc.

Our work on Next Step has forced us to examine the ethical issues relevant at each phase and to realise what a balancing act it could be to, on the one hand, ensure the privacy of community members while, on the other hand, also protecting them and ensuring their safety and fulfilling the government's legal responsibilities. The work has highlighted the complexity of the problem, especially when an online community involves a government department and welfare recipients with a dependency relationship with that department. We believe that special care must be taken in such cases and hope that governments wishing to engage with their citizens using social media will pay attention to these concerns.

## ACKNOWLEDGEMENTS

This research has been funded under the Human Services Delivery Research Alliance (HSDRA) between the CSIRO and the Australian Department of Human Services. We would like to thank Payam Aghaei Pour, Hon Hwang, Brian Jin, Alex Sun and Bo Yan at CSIRO for their contribution to the implementation of this work, the Digital Media team in the Communication Division at the Australian Government's Department of Human Services for their contribution and support, and all the participants in Next Step.